\documentclass[12pt]{article}
\usepackage{amsmath,epsfig,graphicx,amssymb}
\usepackage[margin=0.7in]{geometry}
\newcommand{\beq}{\begin{equation}}
\newcommand{\eeq}{\end{equation}}
\newcommand{\ben}{\begin{eqnarray}}
\newcommand{\een}{\end{eqnarray}}
\newcommand{\bes}{\begin{subequations}}
\newcommand{\ees}{\end{subequations}}
\newcommand{\bFig}{\begin{figure}}
\newcommand{\eFig}{\end{figure}}

\date{}
\begin{document}
\title{Continuous Quantum-Classical Transitions and Measurement: A Relook}
\author{Partha Ghose\footnote{partha.ghose@gmail.com} \\
The National Academy of Sciences, India,\\ 5 Lajpatrai Road, Allahabad 211002, India}
\maketitle
\begin{abstract}
The measurement problem in quantum mechanics originates in the inability of the Schr\"{o}dinger equation to predict definite outcomes of measurements. This is due to the lack of objectivity of the eigenstates of the measuring apparatus. Such objectivity can be achieved if a unified realist conceptual framework can be formulated in terms of wave functions and operators acting on them for both the quantum and classical domains. Such a framework was proposed and an equation for the wave function written down (\cite{ghose1,ghose2}) which smoothly interpolates between the quantum and classical limits. The arguments leading to the equation are clarified in this paper, and the theory is developed further. The measurement problem in quantum mechanics is then briefly reviewed and re-examined from the point of view of this theory, and it is shown how the classical limit of the wave function of the measuring apparatus leads to a natural solution of the problem of definite measurement outcomes without the need for either collapse or pragmatic thermodynamic arguments. This is consistent with Bohr's emphasis on the primacy of classical concepts and classical measuring devices. Possible tests of the theory using low-dimensional systems such as quantum dots are indicated.
\end{abstract}
\section{Introduction}
Quantum mechanics has been plagued by conceptual difficulties right from its inception. Arguably the most important one of these is the measurement problem originating in the inability of the Schr\"{o}dinger equation to predict definite outcomes of measurements. A definite outcome of a `measurement' of a `micro system' in quantum mechanics requires a {\em non-quantum mechanical} observer, like a `macroscopic classical apparatus' or a `conscious observer', and therefore a ``shifty split of the world into `system' and `apparatus''' that lacks conceptual clarity in the absence of any parameter in the equation to distinguish between quantum and classical states \cite{bell}. Some additional hypotheses have been suggested to `solve' the measurement problem, such as the splitting of the universe into branches \cite{everett}, the existence of hidden trajectories (variables) in addition to the wave function \cite{bohm, bohm2}, spontaneous collapse \cite{spon}, etc. However, no universally accepted solution exists \cite{snap}. It is also not clear in which sense, if any, this `incompleteness' is related to that emphasized by Einstein, Podolsky and Rosen \cite{epr}. 

There is a huge conceptual disconnect between a quantum mechanical and a conventional classical description of a physical system. A multi-particle system in classical mechanics is described by real particles moving in real four-dimensional space-time, whereas a multi-particle system in quantum mechanics is described by complex wave functions in a multidimensional configuration space and spanning a Hilbert space. It is only in the case of a single particle that the configuration space is identical with real space. The ontology of classical physics is that of `realism', whereas that of standard quantum mechanics is a debatable issue \cite{leifer,qbism}. What is required therefore for conceptual clarity is a single equation with a realist ontology that interpolates smoothly between the quantum and classical descriptions. Since quantum mechanics has been so immensely successful in describing the microscopic entities that constitute classical systems like measuring apparatus, it would seem reasonable first to attempt to formulate a similar theory of classical mechanical systems based on a Hilbert space of states and operators acting on the states. If that can be done, one would be in a better position to explore the possibility of the required equation. Complex, square integrable wave functions $\psi(q,p,t)$ for classical statistical mechanical systems were, in fact, introduced way back in 1931-32 by Koopman \cite{K} and von Neumann \cite{vN} for this very reason \cite{wilc}. The dynamical variables $p$ and $q$ are represented in this theory by commuting operators. They used the Liouville equation for $\psi(q,p,t)$ and $\psi^*(q,p,t)$ and derived the classical Liouville equation for the density $\psi^*\psi$ in phase space. In this formulation the amplitude and phase of the classical wave function are decoupled and interference effects are avoided.  However, a smooth quantum-classical connection is not easy to see in this theory because the underlying dynamics is different in the two cases.

Here we propose to use a different equation of motion for the classical wave function \cite{ghose1,ghose2} which is closer in form to the Schr\"{o}dinger equation than the Liouville equation. To see how one can arrive at such an equation, let us first consider the Schr\"{o}dinger equation
\ben 
i\,\hbar\,\frac{\partial \psi_q}{\partial t} &=& \left(-\frac{\hbar^2}{2 m}\, \nabla^2  + V\right)\psi_q\label{Sch}
\een
for a quantum mechanical wave function $\psi_q$.
Writing $\psi_q$ in the polar form $\psi_q = R_q {\rm exp}(iS_q/\hbar)$ and separating the real and imaginary parts, one gets the two coupled equations
\ben
\partial S_q/\partial t + \frac{(\nabla S_q)^2}{2m} + V + Q_q &=& 0,\,\,\,\, Q_q = -\frac{\hbar^2}{2 m}\,\frac{\nabla^2 R_q}{R_q}, \\
\frac{\partial R_q^2}{\partial t} +
\vec{\nabla}\,.\, \left(\,R_q^2\, \frac{\vec{\nabla}S_q}{m}\right) &=& 0.
\een
It is obvious from these equations that apart from the term $Q_q$ (the quantum potential), the first equation is identical with the classical Hamilton-Jacobi equation for the action $S_{cl}$. Further, the second equation, which is the continuity equation for the probability density $R^2_q$, shows that if $R_q$ is chosen to be real at some initial time, it would remain so for all future times. In the absence of the term $Q_q$ these two equations would therefore be decoupled, implying that the amplitude $R_{cl}$ and the phase $S_{cl}/\hbar$ of a classical wave function $\psi_{cl}$ evolve independent of each other, ensuring the absence of observable interference effects. Thus, in the absence of the term $Q_q$ the two equations would be completely equivalent to the classical statistical mechanics of a system of particles with a distribution function $R_{cl}^2$ obeying a continuity equation. 

Notice that
\beq
Q_qR_q = -\frac{\hbar^2\nabla^2 R_q}{2 m}  = \frac{\hat{p}^2}{2m}R_q,\,\,\,\, \hat{p} = -i\hbar \nabla, \label{phat}
\eeq
showing that the quantum potential $Q_q$ is the quantum mechanical kinetic energy and is at the origin of the commutation relation $\left[\hat{p}, q\right] = -i\hbar$. Hence, the term $Q_q$ encapsulates in it {\em all} quantum mechanical effects.

The task therefore is to write an additional term in the Schr\"{o}dinger equation that will exactly cancel the quantum potential. With this in view, consider the equation
\ben 
i\,\hbar\,\frac{\partial \psi_{cl}}{\partial t} &=& \left(-\frac{\hbar^2}{2 m}\, \nabla^2  + V - Q_{cl}\right)\psi_{cl},\,\,\,\, Q_{cl} = -\frac{\hbar^2}{2 m}\,\frac{\nabla^2 R_{cl}}{R_{cl}}.\label{NE}
\een
Writing $\psi_{cl}$ in the polar form $R_{cl}{\rm exp}(iS_{cl}/\hbar)$ and separating the real and imaginary parts, one gets
\ben
\partial S_{cl}/\partial t + \frac{(\nabla S_{cl})^2}{2m} + V &=& 0,\label{1} \\
\frac{\partial R_{cl}^2}{\partial t} +
\vec{\nabla}\,.\, \left(\,R_{cl}^2\, \frac{\vec{\nabla}S_{cl}}{m}\right) &=& 0.\label{2}
\een
These are precisely the equations for the action and particle distribution of classical statistical mechanical systems, as we have seen. Hence, eqn.(\ref{NE}) is the required equation for the classical wave function of such systems. Although this equation involves $\hbar$, notice that $\hbar$ drops out of the eqns (\ref{1}) and (\ref{2}). Hence, formulated in terms of a complex wave function satisfying eqn.(\ref{NE}), classical mechanics does not require $\hbar$ to be zero or small. In semi-classical approximations of quantum mechanics one has to assume that $\hbar$ is small compared to the classical action, write a power series expansion of the action $S$ in powers of $\hbar$ and retain only terms of the order of $\hbar$. Such approximations hold only when the de Broglie wavelength of the particle is small compared to the region of the particle's motion. No such approximations are needed in writing the above equations which are exact. The complete quenching of the quantum potential $Q_{cl}$ is achieved by a term of non-quantum mechanical nature in eqn.(\ref{NE}) and $\psi_{cl}$ is a classical wave function, not a semi-classical approximation of quantum mechanics.
 
Consider now the equation
\ben 
i\,\hbar\,\frac{\partial \psi}{\partial t} &=&
\left(-\frac{\hbar^2}{2 m}\, \nabla^2  + V  - \lambda Q \,\right)\,\psi,\,\,\,\, Q = -\frac{\hbar^2}{2 m}\,\frac{\nabla^2 R}{R}\label{new}
\een
where $\lambda$ is a dimensionless scalar parameter. It reduces to eqn.(\ref{Sch}) for $\lambda = 0$ and to eqn.(\ref{NE}) for $\lambda = 1$. Hence, $\lambda$ is the required parameter in the equation to distinguish between quantum and classical states. 

These two limits can be smoothly connected provided one is prepared to admit the existence of novel physical states characterized by $0 < \lambda < 1$. Such states would be neither quantum mechanical nor classical, being intermediate between them. Hence, we would like to call them `mesostates'. Should such states exist, they would provide the missing connection between the two ostensibly disconnected worlds of quantum and classical mechanics. 

Writing $\psi$ in the polar form, one gets
\ben
\partial S/\partial t + \frac{(\nabla S)^2}{2m} + V + (1 - \lambda)Q &=& 0,\label{3} \\
\frac{\partial R^2}{\partial t} +
\vec{\nabla}\,.\, \left(\,R^2\, \frac{\vec{\nabla}S}{m}\right) &=& 0.\label{4}
\een
Notice that the Schr\"{o}dinger equation can be derived from eqn.(\ref{new}) as a limiting case, but not {\em vice versa}. Hence, eqn.(\ref{new}) is a new interpolating equation that has to be {\em postulated}. Furthermore, notice that the effective quantum potential in the theory is $(1 - \lambda)Q$. Hence, in view of the relation (\ref{phat}), $\hat{p}(\lambda) = \sqrt{1 - \lambda}\hat{p}$, and the commutation relation between the momentum and position operators is
\beq
\left[\hat{p}(\lambda), q\right] = -i\sqrt{1 - \lambda}\hbar,
\eeq
which reduces to the quantum mechanical result in the limit $\lambda = 0$. In the classical limit $\lambda = 1$, however, the momentum and position operators commute {\em without requiring} $\hbar = 0$. 
\section{Density Matrices}
It is essential to use density matrices rather than wave functions in taking the classical limit. Let us first consider the quantum mechanical limit $\lambda = 0$. Writing
\ben
\psi_q &=& a \psi_{q1} + b \psi_{q2},\,\,\,\,\,\,a^2 + b^2 = 1,\label{sup}\\
\psi_{q1} &=& R_{q1} e^{\frac{i}{\hbar}S_{q1}}|u\rangle= R_{q1} e^{\frac{i}{\hbar}S_{q1}}\left(\begin{array}{c}
 										1\\
 										0		
 										\end{array}
 									\right),\\
\psi_{q2} &=& R_{q2} e^{\frac{i}{\hbar}S_{q2}}|v\rangle= R_{q2} e^{\frac{i}{\hbar}S_{q2}}\left(\begin{array}{c}
 										0\\
 										1		
 										\end{array}
 									\right),
\een
where both $\psi_{q1}$ and $\psi_{q2}$ satisfy the Schr\"{o}dinger equation,
one gets the density matrix 
\ben
\rho_q &=& a^2|u\rangle\langle u| + b^2|v\rangle\langle v| + ab\left[|u\rangle\langle v|e^{\frac{i}{\hbar}(S_{q1} - S_{q2})} + |v\rangle\langle u|e^{-\frac{i}{\hbar}(S_{q1} - S_{q2})} \right]\nonumber\\
	&=&\left(\begin{array}{cc}
 										a^2&	 ab\,e^{-i\delta}\\
 										ab\,e^{i\delta} &	b^2		
 										\end{array}
 									\right)		\label{rho}					
\een
where $\delta = (S_{q2} - S_{q1})/\hbar$. Notice that this matrix is singular and corresponds to a single wave function $\psi_q$. 

It is clear from eqn.(\ref{3}) that a system retains a degree of quantum coherence in the domain $0\leq \lambda <1$. Hence, in this domain one can write a superposition of wave functions
\ben
\psi &=& a \psi_1 + b \psi_2,\,\,\,\,\,\,a^2 + b^2 = 1,\label{sup2}\\
\psi_1 &=& R_1 e^{\frac{i}{\hbar}S_1}|u\rangle,\\
\psi_2 &=& R_2 e^{\frac{i}{\hbar}S_2}|v\rangle
\een
where both $\psi_1$ and $\psi_2$ are solutions of equation (\ref{new}). The density matrix is 
\ben
\rho^\lambda &=& a^2 |u\rangle\langle u| + b^2 |v\rangle\langle v| + ab\left[|u\rangle\langle v|e^{-if(\lambda)\delta} + |v\rangle\langle u|e^{if(\lambda)\delta}\right]\nonumber\\
&=&\left(\begin{array}{cc}
 										a^2&	 ab\,e^{-if(\lambda)\delta}\\
 										 ab\,e^{if(\lambda)\delta} &	b^2		
 										\end{array}
 									\right),	
\een
where $\delta = (S_2 - S_1)/\hbar$, and $f(\lambda)$ is a function that goes to unity in the limit $\lambda = 0$. This can be verified for cases where the potential $V$ is time independent by using the relation $\partial S/\partial t = - E$ and introducing a realist ontology through the `guidance condition' 
\beq
p(\lambda) = \nabla S 
\eeq
to write eqn.(\ref{3}) in the form
\ben
p^2(\lambda) &=& 2m\left[E - (V + Q) + \lambda Q \right]\label{plambda}\\
&=& 2m\left[E - (V + Q) \right]\left[1 + \frac{\lambda Q }{E - (V + Q)} \right]\nonumber\\
&\equiv& p_q^2f^2(\lambda)\label{p}
\een
where $p_q$ is the quantum mechanical momentum in the Bohmian sense, and
\beq
S = \int p(\lambda) dx = \int p_q f(\lambda) dx.
\eeq
Clearly, $f(0) = 1$.
In the classical limit $\lambda = 1$, we get
\beq
p^2_{cl} = 2m\left[E - V\right]  \label{pclass}
\eeq 
which is independent of $Q$, and also $S_{cl}$ is independent of $R_{cl}$, as we have seen before. Hence, all coherence is lost, and one cannot write a superposition like (\ref{sup2}) in the classical limit. However, one can still write a density matrix in the diagonal form
\beq
\rho_{cl} = \left(\begin{array}{cc}
 										a^2&	 0\\
 										0 &	b^2		
 										\end{array}
 									\right)
\eeq
which describes a statistical mixure of the classical states $\psi_{1cl}$ and $\psi_{2cl}$.

Notice that this diagonal form of the density matrix describes a true mixture of classical states (described by wave functions) that cannot be obtained from standard decoherence theory \cite{om,joos, zur} which holds only in the quantum mechanical limit $\lambda = 0$ in which the states $\psi_{1q}$ and $\psi_{2q}$ remain quantum mechanical although their mutual coherence is lost. Thus, the diagonal form of the density matrix does not by itself imply classicality in its full sense.

\section{The Measurement Problem}
In any operational theory a system is first prepared in a certain state $|S\rangle$ and allowed to propagate and then a measurement is made on it to determine how the state has evolved. In quantum mechanics this state $|S\rangle$ is made to interact with a macroscopic measuring apparatus $|A\rangle$, also treated quantum mechanically, and this in general produces via unitary Schr\"{o}dinger evolution an entangled state of the system and measuring device,
\beq
|SA\rangle = \sum_{i,j}c_{ij}|S\rangle_i|A\rangle_j\label{ent}
\eeq
where the apparatus states $|A\rangle_j$ are eigenstates of some observable in the measurement basis. Its density operator is
\beq
\rho = |SA\rangle\langle SA|
\eeq
which satisfies the condition $\rho^2 = \rho$, ${\rm Tr} \rho = 1$. This contains off-diagonal interference terms in the presence of which it is not legitimate to conclude that the states $|A\rangle_j$
have definite eigenvalues of the observable. Hence a second measurement has to be performed on the apparatus states, leading again to similar conclusions, requiring a third measurement, and this chain is obviously non-terminating. It has been argued that this infinite regress eventually culminates in a non-quantum mechanical {\em subjective} perception \cite{vN2}. von Neumann represented this termination by a non-unitary projection which produces a reduced density operator
\beq
\hat{\rho} = \sum_i |SA\rangle|A\rangle_{ii}\langle A|\langle SA|
\eeq
with the properties $\hat{\rho}^2 \neq \hat{\rho},\, {\rm Tr}\hat{\rho} = 1$, whose diagonal elements are $|c_{ii}|^2$ and whose off-diagonal interference terms have been deleted \cite{vN}.  

To have an {\em objective} interpretation of the measuring process, as some would prefer \cite{dlp}, it is necessary to truncate von Neumann's chain immediately after the first measurement. This can be done if measurement is regarded as a process in which a micro system interacts with an objective and {\em effectively} classical measuring apparatus with an enormously large number of internal {\em thermodynamic} degrees of freedom which are inevitably excited \cite{wheeler} and act like a heat bath, and which responds by changing its original metastable state to a final thermodynamic equilibrium state without substantially affecting the micro system. After the measurement interaction the wave function of the total system splits into several non-overlapping branches in the pointer basis, and the probability of future overlaps among these branches is rendered negligibly small for all practical purposes (FAPP) due to thermodynamic irreversibility. This is the precursor of modern decoherence theory. 

Neverthless, without a realist ontology in which the micro system (the `particle') enters only one of the branches of the final wave function at a time, no definite measurement outcome is predicted, and the measurement problem remains. Bohmian mechanics \cite{bohm3} goes a step further by introducing precisely this realist ontology, and replacing von Neumann's projection by the `guidance condition'. However, the guidance condition does not lead to irreversibility, and hence even in the Bohmian interpretation the future overlaps among the different branches of the final wave function have to be made highly improbable by invoking thermodynamic arguments \cite{bohm,bohm2,dur}. Hence, only pragmatic (FAPP) solutions to the measurement problem are possible within the objective approach, not a fundamental one. 

As is well known, Bohr emphasized the fundamental role of classical concepts and classical measuring devices \cite{bohr}. Though there is growing experimental evidence in support of superpositions of macroscopically distinct states which have no classical counterparts \cite{macro}, Bohr's position is not negated by them because such states cannot obviously be used as measuring devices for the reasons stated above. The fact of the matter is that there is overwhelming empirical evidence that macroscopically distinct classical states do exist to produce definite outcomes of measurements. It is therefore of paramount importance to explore how the established quantum and classical worlds can be mathematically connected in a fundamental way without having to invoke either subjective interventions or pragmatic thermodynamic arguments. This is what is attempted in this paper.
  
\section{Entanglement}
Let us for simplicity consider a two-particle system described by the equation
\ben
i\,\hbar\,\frac{\partial \psi(\vec{x}_1, \vec{x}_2,t)}{\partial t} &=&
\left(-\frac{\hbar^2}{2 m_1}\, \nabla_1^2 -\frac{\hbar^2}{2 m_2}\, \nabla_2^2 + V(\vec{x}_1, \vec{x}_2) - \lambda_1 Q_1 - \lambda_2 Q_2\right)\psi(\vec{x}_1, \vec{x}_2,t), \label{sch2}\\
Q_1 &=& -\frac{\hbar^2}{2m_1}\frac{\nabla_1^2 R}{R}, \,\,\,\,Q_2 =-\frac{\hbar^2}{2m_2}\frac{\nabla_2^2 R}{R},
\een
with $0 \leq \lambda_{1,2}\leq 1$.
Writing
\beq
\psi(\vec{x}_1, \vec{x}_2,t) = R(\vec{x}_1, \vec{x}_2,t) {\rm exp}\left[\frac{i}{\hbar}S(\vec{x}_1, \vec{x}_2,t)\right].\label{wf2}
\eeq
substituting this in (\ref{sch2}) and separating the real and imaginary parts, we get
\ben
&&\partial S/\partial t + \frac{(\nabla_1 S)^2}{2m_1} + \frac{(\nabla_2 S)^2}{2m_2} + V(\vec{x}_1, \vec{x}_2) + (1 - \lambda_1)Q_1 + (1 - \lambda_2)Q_2 = 0,\label{10}\\
&&\frac{\partial R^2}{\partial t} +
{\bf \nabla}_1\,.\, \left(\,R^2\, \frac{{\bf \nabla}_1\,S}{m_1}\right) + {\bf \nabla}_2\,.\, \left(\,R^2\, \frac{{\bf \nabla}_2\,S}{m_2}\right) = 0. 
\een

Now, for simplicity, consider an entangled pair of qubit systems
\beq
\psi(\vec{x}_1, \vec{x}_2,t) = a\psi_{1+}(\vec{x}_1,t)\psi_{2-}(\vec{x}_2,t) + b\psi_{1-}(\vec{x}_1,t)\psi_{2+}(\vec{x}_2,t),\,\,\,\,a^2 + b^2 = 1,\label{entwf}
\eeq
where $\pm$ denote the two orthogonal components of each system. Writing
\ben
\psi_{i\pm}(\vec{x}_i,t) &=& R_{i\pm}(\vec{x}_i,t)|u_{i\pm}\rangle{\rm exp}\left[\frac{i}{\hbar}S_{i\pm}(\vec{x}_i,t)\right],\,\,\,\,i= 1,2,\nonumber\\
u_{i+}&=& \left(\begin{array}{c}
 										1\\
 										0		
 										\end{array}
 									\right),\,\,\,\,u_{i-}=\left(\begin{array}{c}
 										0\\
 										1		
 										\end{array}
 									\right),
\een
we get
\ben
R^2(\vec{x}_1, \vec{x}_2,t) &=& |a|^2R_{1+}^2R_{2-}^2 + |b|^2R_{1-}^2 R_{2+}^2 + 2 ab R_{1+}R_{1-}R_{2+}R_{2-}{\rm cos}\left[\delta_1 + \delta_2 \right],\nonumber\\
\delta_1 &=& \frac{1}{\hbar}\left(S_{1+}- S_{1-}\right),\,\,\,\,\delta_2 =\frac{1}{\hbar}\left(S_{2+}- S_{2-}\right)  
\een
and
\ben
&&S(\vec{x}_1, \vec{x}_2,t) = \hbar\,{\rm arctan}\theta\nonumber\\
&&\theta =\frac{aR_{1+}R_{2-}{\rm sin}(\frac{1}{\hbar}(S_{1+} + S_{2-})) +bR_{1-}R_{2+}{\rm sin}(\frac{1}{\hbar}(S_{1-} + S_{2+}))}{aR_{1+}R_{2-}{\rm cos}(\frac{1}{\hbar}(S_{1+} + S_{2-})) +bR_{1-}R_{2+}{\rm cos}(\frac{1}{\hbar}(S_{1-} + S_{2+}))}.
\een
The density matrix corresponding to the wave function (\ref{entwf}) is given by
\ben
\rho &=& \left[a^2 |u_{1+}\rangle\langle u_{1+}| + b^2 |u_{1=}\rangle\langle u_{1-}|\right]\otimes 
\left[b^2 |u_{2+}\rangle\langle u_{2+}| + a^2 |u_{2-}\rangle\langle u_{2-}|\right]\nonumber\\ &+& \left[ab |u_{1+}\rangle\langle u_{1-}|e^{i\delta_1} +  ab|u_{1-}\rangle\langle u_{1+}|e^{-i\delta_1}\right]\otimes \left[ab |u_{2+}\rangle\langle u_{2-}|e^{i\delta_2} +  ab|u_{2-}\rangle\langle u_{2+}|e^{-i\delta_2}\right]\nonumber\\
&=& \left(\begin{array}{cc}
 										a^2&	 0\\
 										0 &	b^2		
 										\end{array}
 									\right)_1\otimes \left(\begin{array}{cc}
 										b^2&	 0\\
 										0 &	a^2		
 										\end{array}
 									\right)_2\nonumber\\ &+& \left(\begin{array}{cc}
 										0&	 ab\,e^{if_1(\lambda_1)\delta_1}\\
 										ab\,e^{-f_1(\lambda_1)i\delta_1} &	0	
 										\end{array}
 									\right)_1\otimes \left(\begin{array}{cc}
 										0&	 ab\,e^{if_2(\lambda_2)\delta_2}\\
 										ab\,e^{-if_2(\lambda_2)\delta_2} &	0		
 										\end{array}
 									\right)_2 ,\label{den1}
\een
where $f_{1,2}(\lambda_{1,2})$ are defined analogous to eqn.(\ref{p}). 

Now, let system 1 be quantum mechanical ($\lambda_1 = 0$) and system 2 be a classical apparatus ($\lambda_2 = 1$). Then there is no observable phase relation between the two apparatus states which are therefore incoherent, and the density matrix is given by
\beq
\hat{\rho} =\left(\begin{array}{cc}
 										a^2&	 0\\
 										0 &	b^2		
 										\end{array}
 									\right)_1\otimes \left(\begin{array}{cc}
 										b^2&	0\\
 										0 &	a^2		
 										\end{array}
 									\right)_2 + \left(\begin{array}{cc}
 										0&	 ab\,e^{i\delta_1}\\
 										ab\,e^{-i\delta_1} &	0		
 										\end{array}
 									\right)_1\otimes\left(\begin{array}{cc}
 										0&	 0\\
 										0 &	0		
 										\end{array}
 									\right)_2  
\eeq
which indicates a completely mixed state {\em in spite of system 1 remaining coherent and quantum mechanical}. Note that without a wave function description of the classical apparatus states, one could not have derived this result.

\section{An Example of Measurement}
Let us take the case of an observation designed to measure some observable $\mathcal{Q}$ of a particle with wave function $\psi_1(\vec{x},t)$. Let the wave function of the apparatus be $\psi_2(y,t)$ where $y$ is the relevant coordinate of the apparatus. The interaction Hamiltonian is taken to be
\beq
H_I = -g\mathcal{Q}p_y
\eeq
where $g$ is a suitable coupling parameter and $p_y$ is the momentum corresponding to $y$. During the impulsive interaction the free evolution of the two systems can be ignored, and
\ben
i\hbar\frac{\partial \Psi}{\partial t} &=& -g\mathcal{Q}p_y\Psi = \left(\frac{ig}{\hbar}\right)\mathcal{Q}\frac{\partial \Psi}{\partial y}
\een
is a good approximation, and where, under the assumption of discrete eigenfunctions, 
\ben
\Psi(\vec{x},y,t) &=& \sum_q \psi_{1q}(\vec{x})\psi_{2q}(y,t),\\
\mathcal{Q}\psi_{1q}(\vec{x}) &=& q\psi_{1q}(\vec{x}).
\een
Initially the electron and the apparatus are independent, and hence
\beq
\Psi(\vec{x},y) = \psi_{10}(\vec{x})\psi_{20}(y) = \psi_{20}(y)\sum_q c_q\psi_{1q}(\vec{x}).
\eeq
The final wave function is given by
\beq
\Psi(\vec{x},y,t) = \sum_q c_q\psi_{1q}(\vec{x})\psi_{20}(y - gqt/\hbar^2).\label{final}
\eeq
which is an entangled wave function embodying a correlation between the eigenvalue $q$ of $\mathcal{Q}$ and the apparatus coordinate $y$. Writing
\beq
\Psi(\vec{x},y,t) = R(\vec{x},y,t){\rm exp}\left[\frac{i}{\hbar}S(\vec{x},y,t)\right],\label{psi2}
\eeq
we get
\ben
&&\partial S/\partial t + \frac{(\nabla_x S)^2}{2m_1} + \frac{(\partial_y S)^2}{2m_2} + (1 - \lambda_1)Q_1 + (1 - \lambda_2)Q_2 = 0,\label{10a}\\
&&Q_1 = -\frac{\hbar^2}{2m_1}\frac{\nabla_x^2 R}{R}, \,\,\,\,Q_2 =-\frac{\hbar^2}{2m_2}\frac{\partial_y^2 R}{R},\\
&&\frac{\partial R^2}{\partial t} +
{\bf \nabla}_x\,.\, \left(\,R^2\, \frac{{\bf \nabla}_x\,S}{m_1}\right) + \partial_y \left(\,R^2\, \frac{\partial_y S}{m_2}\right) = 0. 
\een
If the particle is quantum mechanical ($\lambda_1 = 0$), eqn. (\ref{10a}) must be replaced by
\beq
\partial S/\partial t + \frac{(\nabla_x S)^2}{2m_1} + \frac{(\partial_y S)^2}{2m_2} + Q_1 + (1 - \lambda_2)Q_2 = 0.
\eeq
Using the guidance condition, one can write the velocities as
\ben
\frac{d\vec{x}_1}{dt} &=& \frac{1}{m_1}\vec{\nabla}_x S(\vec{x},y,t),\\
\frac{dy}{dt} &=& \frac{1}{m_2}\partial_y S(\vec{x},y,t),
\een 
and the second order equations of motion are
\ben
\frac{d\vec{p}_1}{dt} &=& -\vec{\nabla}_x Q_1,\\
\frac{dp_y}{dt} &=& - \partial_y (1 - \lambda_2)Q_2.
\een
 
A specific example would be spin measurement with $\mathcal{Q} = \sigma_z$. In this case the apparatus is a dipole magnetic field in the $xz$ plane with a gradient in the $z$-direction, and let
\ben  
\psi_1(x,z) &=& \left(c_{+}|u_{+}\rangle + c_{-}|u_{-}\rangle \right) f(z)e^{ikx},\\
\sigma_z |u_{\pm}\rangle &=& \pm |u_{\pm}\rangle
\een
be the wave function of a spin-$\frac{1}{2}$ particle travelling in the $x$-direction and passing through the magnet at $t=0$, $f(z)$ being a Gaussian wave packet and $c_{\pm}$ the probability amplitudes for the spin-$\uparrow$ and spin-$\downarrow$ states satisfying the condition $c_{+}^2 + c_{-}^2 = 1$. Then the initial wave function is
\beq
\Psi(x,z) = \psi_{20}(z) \psi_1(x,z), 
\eeq
and the interaction Hamiltonian is
\beq
H_I = i\hbar \mu_B\frac{\partial B}{\partial z}.
\eeq
As a result of the impulsive action of the magnetic field, two wave packets are created which begin to separate in the $z$ direction, and after some time $t$ the wave function has the form
\beq
\Psi(x,z,t) = \left[c_{+}f_{+}(x,z,t)|u_{+}\rangle\psi_{20}(z_{+}) + c_{-}f_{-}(x,z,t)|u_{-}\rangle\psi_{20}(z_{-}) \right],\label{final2}
\eeq 
where $f_{\pm}(x,z,t)$ are the evolved wave packets \cite{dewd1} and $z_{\pm}$ are the corresponding shifted coordinates of the apparatus. This is an entangled state, though the particle is in one of these wave packets only. Now, if the apparatus states are classical ($\lambda_2 = 1$), their relative phases are not observable and they are incoherent. Consequently, the density matrix is given by
\ben
\rho &=& \left(\begin{array}{cc}
 										c_+^2&	 0\\
 										0 &	0		
 										\end{array}
 									\right)_1\otimes \left(\begin{array}{cc}
 										1&	 0\\
 										0 &	0		
 										\end{array}
 									\right)_2 + \left(\begin{array}{cc}
 										0&	 0\\
 										0 &	c_-^2	
 										\end{array}
 									\right)_1\otimes \left(\begin{array}{cc}
 										0&	 0\\
 										0 &	1		
 										\end{array}
 									\right)_2 + \left(\begin{array}{cc}
 										0&	 c_+c_-\,e^{i\delta_1}\\
 										0 &	0		
 										\end{array}
 									\right)_1\otimes \left(\begin{array}{cc}
 										0&	 0\\
 										0 &	0		
 										\end{array}
 									\right)_2\nonumber\\ &+& \left(\begin{array}{cc}
 										0&	 0\\
 										c_+c_-\,e^{-i\delta_1} & 0		
 										\end{array}
 									\right)_1\otimes \left(\begin{array}{cc}
 										0&	 0\\
 										0 &	0		
 										\end{array}
 									\right)_2  
\een
where $\delta_1$ is the phase difference between the two spin states. It is diagonal and shows that after the measurement interaction the system is a statistical mixture of two wave functions $f_{+}(x,z,t)|u_{+}\rangle\psi_{20}(z_{+})$ and $f_{-}(x,z,t)|u_{-}\rangle\psi_{20}(z_{-})$ with the particle in only one of them, resulting in the prediction of definite measurement outcomes with probabilities $c^2_{+}$ and $c^2_{-}$ respectively without collapse, i.e. without one of these wave functions having to vanish every time a definite outcome occurs. The diagonal form of the density matrix is due to the fact that the two apparatus states $\psi_{20}(z_{+})$ and $\psi_{20}(z_{-})$ are incoherent, i.e. have no observable phase relation.

The origin of the debate on nonlocality in quantum mechanics can be traced back to Einstein's observations at the 1927 Solvay Conference. Consider the case of an ontic wavefunction $\psi$ describing a single particle \cite{bac}. After passing through a small hole in a screen, the wave function of the particle spreads out on the other side of it in the form of a spherical wave, and is finally detected by a large hemispherical detector. The wave function propagating towards the detector does not show any privileged direction. Einstein observes:
\begin{quote}
If $|\psi|^2$ were simply regarded as the probability that at a certain point a given particle is found at a given time, it could happen that {\em the same} elementary process produces an action in {\em two or several} places on the screen. But the interpretation, according to which the $|\psi|^2$ expresses the probability that {\em this} particle is found at a given point, assumes an entirely peculiar mechanism of action at a distance, which prevents the wave continuously distributed in space from producing an action in {\em two} places on the screen.
\end{quote}
Einstein later remarks that this `entirely peculiar mechanism of action at a distance' is in contradiction with the postulate of relativity. Hence, the joint assumption of reality of the wave function and locality is incompatible with the Schr\"{o}dinger equation together with the Born rule. We must point out that the nonlocality pointed out by Einstein in this example is clearly a consequence of implicitly assuming a `projective measurement' by the detector, and hence can be eliminated by first considering the detector to be in a `meso' state and then going to the classical limit, as explained above.

\section{The Case of EPR Entanglement}

Consider the EPR-Bohm wave function to have the initial form
\beq
\psi_1(z_a^\prime,z_b) = f_a(z_a^\prime)f_b(z_b)\frac{1}{\sqrt{2}}\left[|u^a_{+}\rangle |v^b_{-}\rangle - |u^a_{-}\rangle |v^b_{+}\rangle \right]
\eeq
where $f_a(z_a^\prime)$ and $f_b(z_b)$ are Gaussian wave packets, $z_a^\prime$ and $z_b$ are coordinates of two particles $a$ and $b$ in the $z^\prime$ and $z$ directions respectively, and $\sigma^a_{z}|u^a_{\pm}\rangle = \pm |u^a_{\pm}\rangle$, $\sigma^b_{z}|v^b_{\pm}\rangle = \pm |v^b_{\pm}\rangle$ \cite{dewd2}. Denoting the initial state of the measuring apparatus by
\beq
\psi_{20}(z^\prime_A,z_B) = \psi_A(z^\prime_A)\psi_B(z_B)
\eeq
where $\psi_A(z^\prime_A)$ and $\psi_B(z_B)$ are the apparatus wave functions used by Alice and Bob respectively, the wave function of the initial state is
\beq
\Psi(z_a^\prime,z_b, z_A^\prime,z_B) = \psi_1(z_a^\prime,z_b)\psi_{20}(z^\prime_A,z_B).
\eeq
Therefore, the entangled state at time $t$ after the impulsive measurement interaction is
\ben
\Psi(z_a^\prime,z_b, z_A^\prime,z_B,t) &=& \frac{1}{\sqrt{2}}f_{a}(z_a^\prime,t)f_{b}(z_b,t)\left[|u^a_{+}\rangle|v^b_{-}\rangle\psi_A(z^\prime_{A+})\psi_B(z_{B-})\right]\nonumber\\ &-& \frac{1}{\sqrt{2}}f_{a}(z_a^\prime,t)f_{b}(z_b,t)\left[|u^a_{-}\rangle|v^b_{+}\rangle\psi_A(z^\prime_{A-})\psi_B(z_{B+})\right]\label{epr} 
\een
where the notation follows that of the previous section with obvious generalizations. The density operator is
\ben
\rho &=& \frac{1}{2}\left[|u^a_{+}\rangle\langle u^a_{+}| |v^b_{-}\rangle\langle v^b_{-}||u^A_{+}\rangle\langle u^A_{+}| |v^B_{-}\rangle\langle v^B_{-}| + |u^a_{-}\rangle\langle u^a_{-}| |v^b_{+}\rangle\langle v^b_{+}||u^A_{-}\rangle\langle u^A_{-}| |v^B_{+}\rangle\langle v^B_{+}|\right]\nonumber\\ &-& \frac{1}{2}\left[e^{i\delta_{ab}} |u^a_{+}\rangle\langle u^a_{-}| |v^b_{-}\rangle\langle v^b_{+}|e^{i\delta_{AB}}|u^A_{+}\rangle\langle u^A_{-}| |v^B_{-}\rangle\langle v^B_{+}|\right]\nonumber\\ &-& \frac{1}{2}\left[e^{-i\delta_{ab}}|u^a_{-}\rangle\langle u^a_{+}| |v^b_{+}\rangle\langle v^b_{-}|e^{-i\delta_{AB}}|u^A_{-}\rangle\langle u^A_{+}| |v^B_{+}\rangle\langle v^B_{-}| \right],
\een
where 
\ben
\delta_{ab} &=& (\delta_{a+} - \delta_{a-}) + (\delta_{b-} - \delta_{b+}),\\
\delta_{AB} &=& (\delta_{A+} - \delta_{A-}) + (\delta_{B-} - \delta_{B+}).
\een
In the realist (Bohmian) interpretation the particles $a$ and $b$ are always in one of their two possible states determined by the wave functions (\ref{epr}), resulting in definite outcomes. 

Now, if the apparatus states are classical, they are incoherent and the density matrix is diagonal: 
\ben
\hat{\rho} &=& \frac{1}{2}\left[|u^a_{+}\rangle\langle u^a_{+}| |v^b_{-}\rangle\langle v^b_{-}||u^A_{+}\rangle\langle u^A_{+}| |v^B_{-}\rangle\langle v^B_{-}| + |u^a_{-}\rangle\langle u^a_{-}| |v^b_{+}\rangle\langle v^b_{+}||u^A_{-}\rangle\langle u^A_{-}| |v^B_{+}\rangle\langle v^B_{+}|\right].
\een 
This does not, however, imply that the EPR-Bohm state collapses into one of the two wave functions in (\ref{epr}) every time a definite outcome occurs. The diagonal form is only indicative of the fact that the relative phases between the classical apparatus states are not observable. Hence, {\em there being no collapse, there is no measurement induced nonlocality in this theory}.

It must be emphasized that the nonlocality inherent in the quantum potential $Q$ is not connected with `measurement' as such, and that it occurs in multi-dimensional configuration spaces.
The EPR paradox is resolved in realist theories of this kind, such as Bohmian mechanics, by showing that the nonlocal and spin-dependent quantum potential associated with the total quantum state rotates the spin vectors of the two particles in such a manner as always to preserve the correlations implied by the quantum state \cite{dewd1,dewd2}.

\section{Significance of $\lambda$}
So far $\lambda$ has played the role of a mathematical parameter which characterizes the various modes in which matter can exist, namely quantum mechanical ($\lambda = 0$), classical ($\lambda = 1$) and perhaps also `mesostates' ($0 <\lambda < 1$), and provides a smooth link between them. However, its physical significance needs to be clarified. It could also be that $\lambda$ is the unit step function, and no mesostates exist. That would rule out smooth quantum-classical transitions. 

Since $0 \leq\lambda\leq 1$, it can be interpreted as a probabilistic measure of the `classicality' of a state, $\lambda = 1$ being $100 \%$ classical and $\lambda = 0$ being $0 \%$ classical. Mesostates would presumably be multi-particle states with varying degrees of `classicality', depending on the number of constituents, their internal structure, temperature, etc., i.e. with varying degrees of the presence of the quantum potential which encapsulates quantum coherence. The quantum-classical transition is therefore a `decoherence' process in which the quantum potential is gradually quenched. It is therefore different from standard environment-induced decoherence which occurs {\em within} quantum mechanics. Hence, $\lambda$ is also a probabilistic measure of decoherence, the maximum occuring for fully classical states with $\lambda = 1$.              

The probabilistic nature of $\lambda$ indicates an underlying stochastic process. Now, the cumulative distribution functions (CDFs) for many common probability distribution functions are sigmoidal. For example, the CDF of a normal distribution,
\beq
F (x) = \int_{-\infty}^x \frac{e^{-t^2/2}}{\sqrt{2\pi}},\label{cdf}
\eeq
is a sigmoid which is the probability that the random variable $t$ takes on a value less than or equal to $x$. Hence, modelling the underlying stochastic process by a single random variable, one can interpret $\lambda$ as the CDF of its probability density function. However, since the integral (\ref{cdf}) does not exist in a simple closed form and is usually computed numerically, for illustrative purposes one can use the logistic function 
\beq
\lambda = \frac{1}{1 + e^{-b(t - t_0)}}\label{log}
\eeq
which is qualitatively similar in form, and where $b$ is a measure of the steepness of the sigmoid and $t_0$ is the $t$-value of the sigmoid's midpoint. It reduces to the $\theta$ function in the limit $b \rightarrow \infty$.
Hence
\beq  
1 - \lambda = \frac{1}{1 + e^{b(t - t_0)}}\label{log2}
\eeq
is the complementary CDF to the logistic function, and it is this function that occurs in eqns.(\ref{3}).

In general, the underlying stochastic process may involve multiple random variables, and hence one must use the joint probability density function or the joint probability mass function depending on whether the random variables are all continuous or discrete.

\section{Particle in a Box}
Let us now consider the problem of a particle in a box which has acquired immense practical importance as a theoretical tool to study quantum dots. As is well known, the solution of the time independent Schr\"{o}dinger equation
\beq
- \frac{\hbar^2}{2m}\frac{d^2 \psi_q}{dx^2} = E\psi_q
\eeq
for a particle inside a one-dimensional box of length $L$ is
\beq
\psi_{qn} = \sqrt{\frac{2}{L}} {\rm sin} k_n x,\,\,\,\,k_n = \frac{n\pi}{L},\,\,n= 1,2,3,\cdots
\eeq
and the quantized energies are given by
\beq
E_n = \frac{n^2\pi^2\hbar^2}{2mL^2}.\label{E1}
\eeq
It will be instructive to see how this result can be derived in Bohmian mechanics. The starting point will be eqn.(\ref{plambda}) with $V=0, \,\lambda =0$ for every wave function $\psi_{qn}$, namely
\beq
p^2_n(0) = 2m\left[E_n - Q_n\right].
\eeq
One can now calculate the quantum potentials
\beq
Q_n = -\frac{\hbar^2}{2m}\frac{|\psi_{qn}|^{\prime\prime}}{|\psi_{qn}|} = \frac{\hbar^2k_n^2}{2m} = \frac{n^2\pi^2 \hbar^2}{2mL^2}
\eeq
where $^\prime$ denotes the derivative w.r.t $x$, and notice that the quantum mechanical result (\ref{E1}) follows provided $p_n(0) = 0$, which shows that a quantum mechanical particle in a box is at rest, and that the quantized energies are simply the quantum potentials.

Let us now see what the time independent version of eqn.(\ref{new}), namely
\beq
- \frac{\hbar^2}{2m}\frac{d^2 \psi}{dx^2} = (E +\lambda Q)\psi \label{new2}
\eeq
has to say about the particle in a box. Our starting point will be eqn.(\ref{plambda}) with $V=0$ for every $n$, i.e.
\ben
p_n^2(\lambda) &=& 2m\left[E_{n\lambda} - (1- \lambda)Q_n\right],
\een
which reduces to the quantum mechanical result for $E_n$ in the limit $\lambda = 0$ provided $p_n(0) = 0$. If $0 <\lambda <1$, we have
\beq
E_{n\lambda} = (1 - \lambda)Q_n = (1 - \lambda)\frac{n^2\pi^2\hbar^2}{2mL^2} = (1 - \lambda)E_n\label{E2}
\eeq
if one sets $p_n(\lambda) = 0$ to avoid quantized momenta.
However, in the classical limit $\lambda =1$, the quantum potentials get cancelled, and one has 
\beq
E = \frac{p^2_{cl}}{2m}, 
\eeq
and there is no quantization and no restriction on the particle to be at rest. 

\section{Possible Experimental Test}
Electrons in low-dimensional nanostructures like quantum dots, wires and wells exhibit quantized energy levels that do not exist in the bulk materials. Structures that are intermediate between the bulk materials and the extreme nano clusters provide a possible testing ground for the smooth quantum-classical transitions implied by eqn.(\ref{new}). For example, the result (\ref{E2}) for a particle in a box implies that as a 3D box is flattened, the quantized energy levels of the particle begin to emerge and attain the full quantum mechanical values in the limit $\lambda = 0$. Conversely, in the bottom-up approach the energy levels should start to disappear. The spacings between the energy levels should change according to the relation 
\beq
\frac{\Delta E_\lambda}{\Delta E} = \frac{E_{(n+1)\lambda} - E_{n\lambda}}{E_{(n+1)} - E_{n}} = (1 - \lambda).
\eeq
Hence, there is a departure from the pure quantum mechanical prediction.
Empirical studies of the changing energy levels as a quantum dot is formed in the bottom-up or top-down method can therefore help test the basic tenet of the theory and also determine the nature of variation of the $\lambda$ parameter by comparisons with predictions based on models such as eqn.(\ref{log2}). In practice, nanostructures contain more than just a single particle, and hence inter-particle effects need to be taken into account. 
\section{Summary}
The basic idea on which the paper has been developed is that empirical evidence strongly suggests the existence of two apparently disparate domains of nature, one macroscopic and classical and the other microscopic and quantum mechanical, with no clear connection between the two. Hence there is a need for some new physics to connect these domains. The fundamental problem with quantum mechanics is its inability to predict definite outcomes of measurements which can only be done with macroscopic classical measuring devices, but there is no scale or parameter in the theory to distinguish between microscopic and macroscopic objects. It is precisely this shortcoming of quantum mechanics that can be overcome by postulating a new equation (\ref{NE}) with a scalar parameter $0\leq \lambda\leq 1$ such that quantum mechanics is recovered in the limit $\lambda =0$ and classical mechanics in the limit $\lambda =1$. The domain $0 <\lambda <1$ describes a new form of matter that is predicted by the theory and is neither classical nor quantum mechanical, being intermediate between them and hence may be called `meso'. The parameter $\lambda$ is a probabilistic measure of the `classicality' of a state. Given a real ontology, this theory predicts definite outcomes of measurements with classical measuring devices, without requiring any collapse and collapse-induced nonlocality.

The theory is testable in principle with nanostructures like the so called quantum dots, quantum wires and quantum wells which are `mesostates' according to this theory. Further work is in progress to elaborate on such tests, as well as in extending the theory to electrodynamics, specially to polarization optics.

\section{Acknowledgement}
The author acknowledges a grant from the National Academy of Sciences, India which enabled this work to be undertaken.

\end{document}